\documentstyle[12pt,worldsci]{article}
\begin{document}
\def\lsim{\mathrel{\lower2.5pt\vbox{\lineskip=0pt\baselineskip=0pt
\hbox{$<$}\hbox{$\sim$}}}}
\def\gsim{\mathrel{\lower2.5pt\vbox{\lineskip=0pt\baselineskip=0pt
\hbox{$>$}\hbox{$\sim$}}}}
\def\gs{SU(2)_{\rm L} \times U(1)_{\rm Y}}
\def\wh{{\cal W}}
\def\bh{{\cal B}}
\def\wtu{\wh^{\mu \nu}}
\def\wtd{\wh_{\mu \nu}}
\def\btu{\bh^{\mu \nu}}
\def\btd{\bh_{\mu \nu}}
\def\cl{{\cal L}}
\def\nll{\cl_{\rm NL}}
\def\ecl{\cl_{\rm EChL}}
\def\fpnl{\cl_{\rm FP}^{\rm NL}}
\def\fpl{\cl_{\rm FP}}
\def\msb{{\overline{\rm MS}}}
\def\mh{M_H}
\vspace*{-1cm}

\begin{flushright}
{\large FTUAM94/32\\
  hep-ph 9412317 }
\end{flushright}
\vspace{5mm}

\title{ THE ELECTROWEAK CHIRAL LAGRANGIAN AS AN EFFECTIVE FIELD THEORY
OF THE STANDARD MODEL WITH A HEAVY HIGGS\footnote{
Invited talk presented by M.J.H. at the "Workshop on Electroweak
Symmetry Breaking", Budapest, July 1994.}}
\author{MARIA J. HERRERO \\
{\em  Departamento de F\'{\i}sica Te\'orica\\
  Universidad Aut\'onoma de Madrid\\
  Cantoblanco,\ \ 28049-- Madrid,\ \ Spain} \\
\vspace{0.3cm}
  and \\
\vspace{0.3cm}
  ESTER RUIZ MORALES \\
{\em  Departamento de F\'{\i}sica Te\'orica\\
  Universidad Aut\'onoma de Madrid\\
  Cantoblanco,\ \ 28049-- Madrid,\ \ Spain} \\ }
\maketitle
\setlength{\baselineskip}{2.6ex}
\begin{abstract}
\noindent

Using effective field theory methods, we integrate out the standard
model Higgs boson to one loop and represent its non-decoupling
effects by a set of gauge invariant effective operators of the
electroweak chiral Lagrangian.  We briefly discuss the relation
between the renormalization of both the standard model and the
effective theory, which is crucial for a correct understanding and
use of the electroweak chiral Lagrangian.  Some examples have been
chosen to show the applicability of this effective Lagrangian
approach in the calculation of low energy observables in electroweak
theory.

\end{abstract}

\section{Introduction}

The original motivation for studying chiral Lagrangians in the
context of electroweak (EW) interactions\cite{DH} was to find an
effective theory that could parametrize at low energies the physics
of the $\gs$ breaking dynamics\cite{FER}. The basic assumption made
in this approach is that whatever the interactions that trigger the
symmetry breaking may be, the particles involved in this breaking
dynamics are heavier than the W and Z bosons. In that case, it is
possible to describe electroweak physics at energies $\approx M_Z$ by
a low energy effective Lagrangian formulated just in terms of the
relevant degrees of freedom\footnote{ The fermionic interactions,
that are ignored in all the discussion, are assumed to be the same as
in the SM.} at low energies, namely the gauge and would-be Godstone
boson fields.  The Higgs field is removed from the physical spectrum
of the theory, either because it supposed to be heavy or because one
wants to describe some other alternative symmetry breaking
scenario.

The lowest order term of this effective lagrangian is a gauged
non-linear sigma model\cite{AB,LON}, that respects the basic
symmetries of the standard model (SM), namely $\gs$ gauge symmetry
spontaneously broken to $U(1)_{\rm em}$ and custodial $SU(2)_{\rm C}$
symmetry of the scalar sector, but does not include a particular
dynamics for the symmetry breaking interactions. This non-linear
Lagrangian  is non-renormalizable in the usual sense, because as one
increases the number of loops in a calculation, new divergent
structures appear of higher and higher dimension.  The
non-renormalizability of the theory is directly related to the fact
that the SM Higgs boson does not decouple at low energies. In fact,
this was the  reasoning followed by Appelquist and Bernard
\cite{AB} and Longhitano\cite{LON} when they made the first
systematic analysis of the leading logarithmic Higgs mass dependent
terms in the SM. Their strategy was to classify the new counterterms
needed to absorb all the new divergences generated in a one
loop calculation with the non-linear sigma model. Then, taking into
account that the Higgs mass acts as a regulator of the linear theory,
they identified these new counterterms with the logarithmic Higgs
mass dependent terms in the SM.

Another important consequence that can be obtained just from
the analysis of the non-linear Lagrangian is that these leading
logarithmic terms are not sufficient to discriminate a heavy Higgs
possibility from an alternative symmetry breaking scenario to which
one requires to respect, at low energies, the same symmetries as the
SM. These logarithmic contributions are a consequence of the general
gauge and custodial symmetry requirements of the low energy structure
of EW interactions and therefore they will be the same irrespective
of the particular choice for the breaking dynamics, with the Higgs
mass being replaced by some alternative physical mass. Thus, if one
wants to reveal the nature of the symmetry breaking from low energy
observables, one has to go beyond the leading logarithmic effects.

In order to discriminate between different symmetry breaking
scenarios from low energy observables with chiral Lagrangians, one
has to use the full machinery of effective field theories.  The first
step to be taken is to construct a low energy effective theory that
can be consistently applied up to one loop, that we call the
electroweak chiral Lagrangian (EChL)\cite{HR2,HR3}.  The EChL
consists in the $\gs$ gauged non-linear sigma model introduced
before, plus the whole set of gauge invariant effective operators up
to dimension 4, that were classified by Longhitano. This effective
theory can be renormalized order by order in the loop expansion, in a
similar way as it is done in  chiral perturbation theory\cite{W,GL1}.
In particular, at one loop order, the new divergences generated by a
one loop calculation with the non-linear sigma model can be absorved
into redefinitions of the effective operators. These effective
operators parametrize the low energy effects of the underlying
fundamental dynamics of the symmetry breaking.

Apart from the logarithmic dependence already mentioned, the
effective theory does not give any further theoretical insight on the
values of the effective operators for every particular choice of the
underlying symmetry breaking dynamics. To get this information, a
second step has to be taken and regard the effective operators as the
result of integrating out the heavy fields of some underlying
fundamental dynamics. In a perturbative approach, the fundamental and
effective theories can be related by doing explicit calculations of
the relevant loop diagrams and by matching the predictions of the
full underlying theory (in which heavy particles are present) and
those of the low energy effective theory (with only light degrees of
freedom) at some reference scale.

In the rest of this talk, we will describe how to find the values of
the effective chiral operators when the underlying theory is the
standard model with a heavy Higgs\cite{HR2,HR3}.  We will calculate
the complete non-decoupling (leading logarithms plus constant
contributions) effects of a heavy SM Higgs boson by integrating out
the Higgs field to one loop  and by matching the SM predictions in
the large $\mh$ limit with the predictions from the electroweak
chiral Lagrangian to one loop order.

Although the non-decoupling effects of the SM Higgs boson have been
extensively analyzed for most of the low energy observables, we
believe that the classification of these effects in terms of the
effective EChL operators may be interesting for several reasons.
First of all, the EChL approach provides a gauge invariant and
systematic way of separating the non-decoupling Higgs boson effects
from the rest of the EW radiative corrections in low energy
processes. On the other hand, the EChL is a general framework in
which one can analyze the low energy effects not only of a heavy
Higgs in the SM, but of more general symmetry breaking dynamics
characterized by the absence of light modes. It is therefore
desirable to have the EChL that parametrizes an SM Higgs as a
fundamental reference model.

In section 2 we give the precise formulation of the EChL and the
one-loop renormalized Green's functions of the effective theory. In
section 3, we discuss the matching porcedure used to relate the
effective theory and the SM.  We will comment briefly in section 4
the renormalization of the SM, that has been taken to be on-shell.
In section 5, we set the matching conditions neccesary to completely
determine the EChL parameters for a heavy Higgs.  In that section, we
also give and comment the values of the effective operators obtained
from the matching.  Section 6 is devoted to make a more detailed
analysis of the on-shell renormalization of the effective theory,
that is necessary to calculate, in section 7, several examples of EW
radiative corrections to low energy observables.

\section{The Electroweak Chiral Lagrangian}

The EChL is the most simple effective theory of EW interactions that
parametri\-zes, at low energies, the effects of symmetry breaking
sectors whose typical mass scale is larger than $M_Z$. It is
formulated just in terms of the "light" gauge and would-be Goldstone
fields, satisfying the basic requirement of $\gs$ gauge invariance
spontaneously broken to $U(1)_{\rm em}$:
\begin{equation}
\ecl = \nll + \sum_{i=0}^{13} \cl_{i}. \label{ECL}
\end{equation}
Its basic structure is a gauged non-linear sigma model $\nll$,
where a non-linear parametri\-zation of the would-be Goldstone
bosons is coupled to the $\gs$ gauge fields
\begin{equation}
\nll  =  \frac{v^2}{4}\; Tr\left[ D_\mu U^\dagger D^\mu U \right]
+ \frac{1}{2}\; Tr\left[ \wtd \wtu + \btd \btu
\right] + \cl_{\rm R_\xi} + \fpnl,
\label{NLL}
\end{equation}
where the bosonic fields have been parametrized as
\begin{eqnarray}
U & \equiv & \displaystyle{\exp\left( {i \;
\frac{\vec{\tau}\cdot\vec{\pi}}{v}}\right)},\;\;\;\;\;
v  = 246 \;{\rm GeV}, \;\;\;\;\; \vec{\pi} = (\pi^1,\pi^2,\pi^3),
\nonumber\\[1mm]
 \wh_\mu & \equiv & \frac{ -i}{2}\;
\vec{W}_\mu \cdot \vec{\tau}, \nonumber\\[1mm]
\bh_\mu & \equiv & \frac{ -i}{2} \; B_\mu \;
\tau^3,\label{FPAR}
\end{eqnarray}
and the covariant derivative and the field strength tensors are
defined as
\begin{eqnarray}
D_\mu U & \equiv & \partial_\mu
U - g \wh_\mu U + g' U \bh_\mu, \nonumber\\[1mm]
\wtd & \equiv &  \partial_\mu \wh_\nu - \partial_\nu \wh_\mu -
g [\wh_\mu, \wh_\nu],\nonumber\\[1mm]
\btd & \equiv &  \partial_\mu \bh_\nu - \partial_\nu \bh_\mu.
\label{FTEN}
\end{eqnarray}
The physical fields are given by
\begin{eqnarray}
W^\pm_\mu & = & \frac{W^1_\mu \mp i W^2_\mu}{\sqrt 2},
\nonumber\\[2mm] Z_\mu & = & c\; W^3_\mu - s\; B_\mu
\nonumber, \\[2mm] A_\mu & = & s\; W^3_\mu + c\; B_\mu,
\end{eqnarray}
where $c = \cos \theta_{\rm w}, \; s = \sin
\theta_{\rm w}$ and the weak
angle is defined by $\tan \theta_{\rm w} = g' / g$.

The second term in Eq.(\ref{ECL}) includes the set of
$\gs$ and CP invariant operators up to dimension four
that were classified by Longhitano\cite{LON}
\footnote{The relation with Longhitano's notation can be
found in our work\cite{HR3}.}:
\begin{equation}
\begin{array}{ll}
\cl_{0}  =  a_0 g'^2 \frac{v^2}{4} \left[ Tr\left( T V_\mu \right)
\right]^2 \hspace{1cm} &
\cl_{7}  =  a_7 Tr\left( V_\mu V^\mu \right) \left[ Tr\left( T V^\nu
\right) \right]^2\\[3mm]
\cl_{1}  =  a_1 \frac{i g g'}{2} B_{\mu\nu}  Tr\left( T \wtu
\right)  &
\cl_{8}  =  a_8  \frac{g^2}{4} \left[ Tr\left( T \wtd \right)
\right]^2 \\[3mm]
\cl_{2}  =  a_2 \frac{i g'}{2} B_{\mu\nu} Tr\left( T [V^\mu,V^\nu ]
\right) \hspace{1cm} &
\cl_{9}  =  a_9  \frac{g}{2} Tr\left( T \wtd \right) Tr\left(
T [V^\mu,V^\nu ] \right) \\[3mm]
\cl_{3}  =  a_3  g Tr\left( \wtd [V^\mu,V^\nu ]\right)
\hspace{1cm} &
\cl_{10}  =  a_{10} \left[ Tr\left( T V_\mu \right) Tr\left( T V_\nu
\right) \right]^2 \\[3mm]
\cl_{4}  =  a_4  \left[ Tr\left( V_\mu V_\nu \right) \right]^2
\hspace{1cm} &
\cl_{11}  =  a_{11} Tr\left( ( D_\mu V^\mu )^2 \right)\\[3mm]
\cl_{5}  =  a_5  \left[ Tr\left( V_\mu V^\mu \right) \right]^2 &
\cl_{12}  =  a_{12} Tr\left( T D_\mu D_\nu V^\nu \right) Tr
\left( T V^\mu \right)\\[3mm]
\cl_{6}  =  a_6 Tr\left( V_\mu V_\nu \right) Tr\left( T V^\mu
\right) Tr\left( T V^\nu \right) \hspace{1cm} &
\cl_{13} =  a_{13} \frac{1}{2} \left[ Tr \left( T D_\mu V_\nu
\right) \right]^2 \label{Li}
\end{array}
\end{equation}
\begin{displaymath}
T \equiv  U \tau^3 U^\dagger, \hspace{2cm} V_\mu
\equiv (D_\mu U) U^\dagger.
\end{displaymath}

We have worked in a generic R$_\xi$ gauge; the gauge fixing term
$\cl_{{\rm R}_\xi}$ and the Faddeev-Popov lagrangian $\fpnl$ in
Eq.(\ref{ECL}) were given in our previous work~\cite{HR2}. We refer
the reader to this work for the detailed formulas and a discussion on
these terms.  It is worth just recalling here that $\fpnl$ does not
coincide with the usual Faddeev-Popov lagrangian of the SM due to the
non-linearity of the would-be Goldstone bosons under infinitesimal
$\gs$ transformations.

It is also important to mention that because of the non-linear
realization of the gauge symmetry some of the couplings in $\nll$
have different Feynman rules than in the SM \cite{HR3}.

As we have mentioned in the introduction, the non-linear sigma model
lagrangian in Eq.(\ref{NLL}) is not renormalizable, as increasing the
number of loops in a calculation implies the appearence of new
divergent structures of higher and higher dimension. However, the
EChL is an effective theory that can be renormalized order by order
in the loop expansion.  In particular, at one loop order, the new
divergences generated by a one loop calculation with $\nll$  can be
absorved into redefinitions of the effective operators given in
Eq.(\ref{Li}).  Therefore, one can obtain finite renormalized Green's
functions if one makes a suitable redefinition of the fields and
parameters of the EChL, among which the chiral parameters $a_i$ must
be included\cite{W,GL1}.  Formally, one defines the renormalized
quantities in the effective theory by the following relations
\begin{center}$
\begin{array}{ll}
B_{\mu}^b  \; = \; \widehat{Z}_B^{1/2} \; B_\mu, \hspace{1cm}& g'^b
\; =\; \widehat{Z}_B^{-1/2}\; ( g' - \widehat{\delta g'} ), \\[2mm]
\vec{W}_\mu^b \; =\;  \widehat{Z}_W^{1/2}\; \vec{W}_\mu, &
 g^b \; =\; \widehat{Z}_W^{-1/2}\; ( g - \widehat{\delta g} ),
\\[2mm]
\vec{\pi}^b \; =\; \widehat{Z}_\Phi^{1/2}\; \vec{\pi}, &
v^b \; = \; \widehat{Z}_\Phi^{1/2}\; ( v -\widehat{ \delta v}),
\\[2mm] \xi_B^b \; = \; \xi_B\; ( 1 + \widehat{\delta \xi}_B),
\hspace{1cm}& \xi_W^b \; = \; \xi_W \;( 1 + \widehat{\delta \xi}_W),
\end{array}$
\begin{equation}
a_i^b \; = \; a_i(\mu) \; + \; \delta a_i , \label{RET}
\end{equation}
\end{center}
where the renormalization constants of the effective theory are
$ \widehat{Z}_i \equiv 1 + \widehat{\delta Z_i}$ and the
superscript b denotes bare quantities. We use the hatted
notation to distinguish counterterms and Green's functions in the
effective theory from the corresponding quantities in the SM.

The 1PI renormalized Green's functions of the effective theory
to one loop will be generically denoted by
\begin{equation}
\widehat{\Gamma}^{\rm R} = \widehat{\Gamma}^{\rm T} +
\widehat{\Gamma}^{\rm C} + \widehat{\Gamma}^{\rm L},
\label{GFE}
\end{equation}
where the superscript R denote renormalized function and the
superscripts T, C and L denote the tree level, counterterm and loop
contributions respectively.  We will discuss in section 6 the
on-shell renormalization of the effective theory, giving explicit
expressions for the counterterms introduced in Eq.(\ref{RET}). For
the moment, in order to discuss the matching procedure, we will treat
the counterterm contributions to the renormalized functions of
Eq.(\ref{GFE}) just at a formal level.

\section{The Matching Procedure}

We would like to focus now our attention on the procedure to obtain
the chiral effective parameters for a heavy Higgs.  As we have
already said, once a particular renormalization scheme has been
chosen to fix the counterterms of the effective theory, the
renormalized $a_i(\mu)$ parameters remain as free parameters that can
not be determined within the framework of the low energy effective
theory.  The values of the renormalized chiral parameters can be
constrained from the experiment, as they are directly related to
different observables in scattering processes
\cite{DH,BDV,DHT,HR1,MAY} and in precision electroweak
measurements\cite{HT,DEH,EH,PES} (see also section 7);
but to have any theoretical insight on their values, one has to
relate the effective theory with a particular underlying fundamental
dynamics of the symmetry breaking.

If the underlying fundamental theory is the standard model with a
heavy Higgs, the chiral parameters can be determined by matching the
predictions of the SM in the large Higgs mass limit and those of the
EChL, at one loop level\cite{HR2,HR3}. By heavy Higgs we mean a Higgs
mass much larger than any external momenta and light particle masses
($ p^2, M_Z^2 \ll \mh^2$) so that one can make a low energy
expansion, but smaller than say 1 TeV, so that a perturbative loop
calculation is reliable.

We will impose here the strongest form of matching\cite{GEO} by
requiring that all renormalized one-light-particle irreducible (1LPI)
Green's functions are the same in both theories at scales $\mu \leq
\mh$. The 1LPI functions are, by definition, the Green's functions
with only light particles in the external legs and whose contributing
graphs cannot be disconnected by cutting a single light particle
line.  This matching condition is equivalent to the equality of the
light particle effective action in the two descriptions.  Some other
forms of matching have been discussed in the literature, by requiring
the equality of the two theories at the level of the physical
scattering amplitudes\cite{DH} or connected Green's
functions\cite{DOM}. These requirements, however, complicate the
calculation unnecesarily while give at the end the same results for
the physical observables.

In order to completely determine the chiral parameters in terms of
the parameters of the SM, it is enough to impose matching conditions
in the two, three, and four-point 1LPI renormalized Green's functions
with external gauge fields.  We have worked in a general
R$_\xi$-gauge to show that the chiral parameters $a_i$ are
$\xi$-independent. We use dimensional regularization and the on-shell
substraction scheme.

The SM Green's functions are non-local; in particular, they depend on
$p / \mh$ through the virtual Higgs propagators. One has to make a
large $M_H$ expansion to represent the virtual Higgs boson effects by
the local effective operators ${\cal L}_i$. In this step, care must
be taken since clearly the operations of making loop integrals and
taking the large $\mh$ limit do not commute. Thus, one must first
regulate the loop integrals by dimensional regularization, then
perform the renormalization with some fixed prescription (on-shell in
our case) and at the end take the large $\mh$ limit, with $\mh$ being
the renormalized Higgs mass.  From the computational point of view,
in the large $\mh$ limit we have neglected  contributions that depend
on $(p/\mh)$ and/or $(M_V/\mh, M_V = M_W, M_Z))$ and vanish when the
formal $\mh \rightarrow \infty$ limit is taken.  An illustrative
example of how to take the large $\mh$ expansion of the loop
integrals can be found in our second work\cite{HR3}.

The matching procedure can be summarized by the following relation
among renormalized 1LPI Green's functions
\begin{equation}
\Gamma^{\rm R}_{\rm SM} (\mu ) \; = \;
\widehat{\Gamma}^{\rm R}_{\rm EChL}
(\mu ) \; , \;\;\;\;\;\;\;\;\;\mu \leq \mh,
\label{match}
\end{equation}
where the large Higgs mass expansion of the SM Green's functions has
to be made as explained above.  This matching condition imposes a
relation between the renormalization of the SM and the
renormalization of the effective theory. We have chosen to
renormalize both theories in the on-shell scheme, so that the
renormalized parameters are the physical masses and coupling
constants. Therefore, the renormalized parameters are taken to be the
same in both theories and the matching conditions will provide
relations between the SM and the EChL counterterms
\footnote{ In some related literature on effective field theories
\cite{GEO,SANTA}, the choice of a mass-independent substraction
prescription ($\msb$) in both theories has also been discussed. In
that case, the matching procedure relates the running
$\msb$-renormalized parameters, that are different in the fundamental
and the effective theories.}.

The matching condition (\ref{match}) represents symbolically a system
of tensorial coupled equations (as many as 1LPI functions for
external gauge fields) with several unknowns, namely the complete set
of parameters $a_i^b$ that we are interested in determining and some
constraints relating the counterterms in both theories. In section 5,
we will give the solution of matching equations for the two, three
and four point Green's functions but before that, we have to set a
renormalization prescription for the standard model.

\section{Renormalization of the Standard Model}

We start by writing down the SM lagrangian
\begin{equation}
\cl_{\rm SM} = (D_\mu \Phi)^\dagger (D^\mu \Phi) + \mu^2
\Phi^\dagger \Phi - \lambda (\Phi^\dagger \Phi)^2 +
\frac{1}{2} Tr \left( \wtd \wtu + \btd \btu \right) +
\cl_{{\rm R}_\xi} + \cl_{\rm FP} ,\\[2mm]
\end{equation}
where
\begin{eqnarray}
\Phi & = & \frac{1}{\sqrt 2}\left(
\begin{array}{c}\phi_1 - i \phi_2 \\
 \sigma + i \chi \end{array}\right), \hspace{2cm}
(\pi_1,\pi_2,\pi_3)  \equiv  (-\phi_2,\phi_1,-\chi) ,
\nonumber\\[2mm]
D_\mu \Phi &  = &  ( \partial_\mu + \frac{1}{2} i g
\vec{W}_\mu\cdot\vec{\tau} +
\frac{1}{2} i g' B_\mu) \Phi .
\end{eqnarray}
$\wtd, \btd$ are defined in Eq.(\ref{FPAR},\ref{FTEN}),
$\cl_{{\rm R}_\xi}$ and $\cl_{\rm FP}$ are the usual R$_{\xi}$
gauge fixing and Faddeev--Popov terms of the standard model.

We rescale the fields and parameters as follows
\begin{equation}
\begin{array}{ll}
B_{\mu}^b  = Z_B^{1/2} B_\mu , \hspace{2cm} &
\vec{W}_\mu^b  =  Z_W^{1/2} \vec{W}_\mu ,\\[2mm]
\Phi^b = Z_\Phi^{1/2} \Phi ,&
v^b  = Z_\Phi^{1/2} ( v - \delta v ) , \\[2mm]
g^b  = Z_W^{-1/2} ( g - \delta g ) , &
g'^b  =  Z_B^{-1/2} ( g' - \delta g' ) ,\\[2mm]
\mu^b  =  Z_\Phi^{-1/2} ( \mu - \delta \mu ) ,&
\lambda^b  =  \lambda (1 - \delta \lambda / \lambda) ,\\[2mm]
\xi_B^b  =  \xi_B ( 1 + \delta \xi_B ) , &
\xi_W^b  =  \xi_W ( 1 + \delta \xi_W ) .
\end{array} \label{REPS}
\end{equation}
where the renormalization constants of the SM are
$ Z_i \equiv 1 + \delta Z_i$ and the superscript b denotes bare
quantities.

We have chosen to renormalize the SM in the on-shell scheme. We
choose the physical masses, $\mh$, $M_W$, $M_Z$ and $g$ as our
renormalized parameters.  The weak mixing angle is defined in terms
of physical quantities, as it is usual in the on-shell scheme
\begin{equation}
\cos^2 \theta_W \equiv \frac{M_W^2}{M_Z^2}.
\end{equation}

The 1LPI renormalized Green's functions in the standard model
to one loop will be generically denoted by
\begin{equation}
\Gamma^{\rm R} = \Gamma^{\rm T} + \Gamma^{\rm C} + \Gamma^{\rm L},
\end{equation}
where one has to consider the tree, counterterm and loop
contributions of all the one light particle irreducible diagrams in
the SM; that is, all the diagrams that cannot be disconnected by
cutting a light (non-Higgs) particle line.
The details on the on-shell SM counterterms and the renormalization
that we have chosen for the tadpole and the scalar self-coupling
$\lambda$\cite{MAW} can be found in our second work\cite{HR3}.

\section{Solution to the Matching Equations}

In this section we present the results of our calculation of the
two, three and four Green's functions for
external gauge fields, giving the set of matching
equations that we have imposed and their solution.
The master equations that summarize the complete
set of matching conditions are the following:
\begin{eqnarray}
\Pi_{ab}^{{\rm T} \mu\nu} + \Pi_{ab}^{{\rm C} \mu\nu} +
\Pi_{ab}^{{\rm L} \mu\nu} & = &
\widehat{\Pi}_{ab}^{{\rm T} \mu\nu} +
\widehat{\Pi}_{ab}^{{\rm C} \mu\nu} +
\widehat{\Pi}_{ab}^{{\rm L} \mu\nu} \label{PM}
\nonumber\\[2mm] V_{abc}^{{\rm T} \lambda\mu\nu} +
V_{abc}^{{\rm C} \lambda\mu\nu} +
V_{abc}^{{\rm L}\lambda\mu\nu} & = &
\widehat{V}_{abc}^{{\rm T} \lambda\mu\nu} +
\widehat{V}_{abc}^{{\rm C}
 \lambda\mu\nu} +\widehat{V}_{abc}^{{\rm L}
\lambda\mu\nu} \nonumber\\[2mm]
 M_{abcd}^{{\rm T} \; \mu \nu \rho \lambda} +
 M_{abcd}^{{\rm C} \; \mu \nu \rho \lambda} +
 M_{abcd}^{{\rm L} \; \mu \nu \rho \lambda} & = &
 \widehat{M}_{abcd}^{{\rm T} \; \mu \nu \rho \lambda} +
 \widehat{M}_{abcd}^{{\rm C} \; \mu \nu \rho \lambda} +
 \widehat{M}_{abcd}^{{\rm L} \; \mu \nu \rho \lambda},
\label{MAMA}
\end{eqnarray}
where $ab$ stand for $WW$, $ZZ$, $\gamma\gamma$ and $\gamma Z$;
$abc$ represent $\gamma WW$ and $ZWW$ and
$abcd$ $= \gamma\gamma WW$, $\gamma ZWW$, $ZZWW$, $WWWW$, $ZZZZ$.

The calculation of the one loop contributions is the most involved
part.  One must include all the 1PI diagrams in the EChL and all the
1LPI diagrams in the SM.  1LPI diagrams are those that cannot be
disconnected by cutting a single light particle line, that is, a
non-Higgs particle line.  One must, in principle, account for all
kind of diagrams with gauge, scalar and ghost fields flowing in the
loops.  However, some simplifications occur.  Firstly, a subset of
the diagrams that have only light particles in it is exactly the same
in both models, and their contribution can be simply dropped out from
both sides of the matching equation (\ref{MAMA}). This is the case,
for instance, of the subset of diagrams whith only gauge particles in
them.  Secondly, we have checked explicitely by analyzing the large
$\mh$ expansion of every diagram that in the case of the three and
four point Green's functions, only those diagrams with just scalar
(Goldstone bosons or Higgs) particles in the loops do contribute with
non-vanishing corrections in the large $\mh$ limit to the matching
equations. In the case of the 2-point functions, however, both pure
scalar and mixed gauge-scalar loops do contribute in the large $M_H$
limit. Finally, among the diagrams with pure scalar loops, there are
some with only Goldstone boson particles. One would expect that these
diagrams give the same contributions in the SM and the EChL, however
they do not. The reason is the already mentioned differences between
the SM and the EChL Feynman rules of some vertices. Therefore, care
must be taken to include these diagrams in both sides of the matching
equations.

In our works\cite{HR2,HR3}, the explicit expressions for the
tree, counterterms and loop contributions to the matching equations
for the different Green's functions can be found.

There is just one compatible solution to the complete set of matching
conditions given by the following values of the bare electroweak
chiral parameters
\begin{eqnarray}
a_0^b &  =  &  \frac{1}{16 \pi^2} \frac{3}{8}
\left( \Delta_\epsilon - \log \frac{\mh^2}{\mu^2} +
\frac{5}{6}\right), \nonumber \\[2mm]
a_1^b & = &  \frac{1}{16 \pi^2} \frac{1}{12}
\left( \Delta_\epsilon - \log \frac{\mh^2}{\mu^2}
+ \frac{5}{6} \right), \nonumber \\[2mm]
a_2^b & = &  \frac{1}{16 \pi^2}
\frac{1}{24} \left( \Delta_\epsilon - \log
\frac{\mh^2}{\mu^2} +
\frac{17}{6} \right), \nonumber \\[2mm]
a_3^b & = & \frac{-1}{16 \pi^2} \frac{1}{24}
\left( \Delta_\epsilon - \log \frac{\mh^2}{\mu^2} +
\frac{17}{6} \right), \nonumber \\[2mm]
a_4^b & = & \frac{-1}{16 \pi^2} \frac{1}{12}
\left( \Delta_\epsilon - \log \frac{\mh^2}{\mu^2} +
\frac{17}{6}\right), \nonumber \\[2mm]
a_5^b & = & \frac{M_W^2}{2 g^2 \mh^2} - \frac{1}{16 \pi^2}
\frac{1}{24} \left( \Delta_\epsilon - \log \frac{\mh^2}{\mu^2}
+ \frac{79}{3} - \frac{ 27 \pi}{2 \sqrt{ 3}} \right),\nonumber\\[2mm]
a_{11}^b  & = &
\frac{-1}{16 \pi^2}\frac{1}{24}, \nonumber \\[2mm]
a_6^b & = & a_7^b \; =\; a_8^b \; = \; a_9^b \; = \; a_{10}^b \; = \;
a_{12}^b \; = \; a_{13}^b \; = \; 0. \label{aMH}
\end{eqnarray}
where
\begin{equation}
\Delta_\epsilon  =  \frac{2}{\epsilon} - \gamma_{\rm E} +
\log 4 \pi, \hspace{1cm} \epsilon = 4 - D
\label{EPS}
\end{equation}
We would like to make some remarks on this result for the chiral
parameters:
\begin{enumerate}
\item First of all, we agree with the $1/\epsilon$ dependence
of the $a_i^b$ parameters that was first calculated by Longhitano
\cite{LON} looking at the divercences of the non-linear sigma model.
We see therefore that the divergences generated with the $\nll$ to
one loop are exactly canceled by the $1/\epsilon$ terms in the
$a_i^b$'s.
\item The values of $a^b_4$ and $a^b_5$ agree with previous
results\cite{DH}, where the equivalence theorem\cite{EQV} was used in
comparing the scattering amplitudes for Goldstone bosons in the SM
\cite{DW} and the EChL. These values, however,
do not coincide with the values\cite{EH} obtained
when only the pure Higgs loops are taken into account.
\item It is important to realize that the matching procedure fixes
completely the values of the bare parameters $a_i^b$ in terms of the
renormalized parameters of the SM, ensuring the equality of
the two theories at low energies.
\item Eqs.(\ref{aMH}) give the complete non-decoupling effects
of a heavy Higgs, that is, the leading logarithmic dependence
on $\mh$ and the next to leading constant contribution
to the electroweak chiral parameters.
The $a_i$'s are accurate up to corrections of the order
$(p/\mh)$ where $p \approx M_Z$ and higher order
corrections in the perturbative expansion.
\item We demonstrate that the $a_i$'s are gauge independent, as
expected.
\item Only one custodial breaking operator, the one corresponding to
$a_0$  which has dimension
2, is generated when integrating out the Higgs at one loop.  No
custodial breaking operator of dimension four is generated.
\item  There is only one effective operator, the one corresponding to
$a_5$, that gets a tree level contribution.
Its expression in terms of renormalized SM parameters
depends on the renormalization prescription that one has
chosen in the standard model, on-shell in our case.
In our second work\cite{HR3}, we discussed how this
effective parameter changes if a different renormalization is chosen
for the SM. For instance, in the $\msb$ scheme, one would obtain
\begin{displaymath}
a^b_5 = \frac{1}{16 \lambda_\msb} -
\frac{1}{16 \pi^2} \frac{1}{24}
\left( \hat{\Delta}_\epsilon - \frac{67}{6}\right).
\end{displaymath}
where $\lambda_\msb$ is the renormalized scalar self-coupling in the
$\msb$.  We would like to emphasize with this example that the bare
chiral parameters for a given underlying theory, the SM in our case,
must be computed once a renormalization prescription of the
underlying theory is chosen. The explicit expression of the chiral
parameters will vary from one prescription to another, but the
numerical value remains the same, and the connection between
different prescriptions  can be clearly and easily established.
\end{enumerate}
{}From the matching conditions, one also obtains the following relations
among the counterterms of the two theories
\begin{eqnarray}
\Delta Z_W & = & \frac{- g^2}{16 \pi^2} \frac{1}{12} \left(
\Delta_\epsilon - \log \frac{\mh^2}{\mu^2} + \frac{5}{6} \right),
\\[2mm]
\Delta Z_B & = & \frac{- g'^2}{16 \pi^2} \frac{1}{12} \left(
\Delta_\epsilon - \log \frac{\mh^2}{\mu^2} + \frac{5}{6} \right),
\nonumber\\[2mm]
\Delta \xi_W & = & \Delta Z_W, \hspace{1cm}
\Delta \xi_B  =  \Delta Z_B, \nonumber \\[2mm]
\frac{\Delta g^2}{g^2} & = & \frac{\Delta g'^2}{g'^2}
= 0, \nonumber \\[2mm]
\Delta Z_\phi - 2 \frac{\Delta v}{v} & = & \frac{g^2}{16 \pi^2}
\left[ - \frac{\mh^2}{8 M_W^2} + \frac{3}{4} \left(
\Delta_\epsilon - \log \frac{\mh^2}{\mu^2} + \frac{5}{6} \right)
\right. \nonumber \\[2mm]
& & \left. + \frac{1}{4} \frac{\xi_Z}{c^2}  \left(
\Delta_\epsilon - \log \frac{\xi_Z M_Z^2}{\mu^2} + 1 \right)
+ \frac{1}{2} \xi_W  \left(
\Delta_\epsilon - \log \frac{\xi_W M_W^2}{\mu^2} + 1 \right)
\right] \nonumber
\end{eqnarray}
where
\begin{equation}
\Delta Q \equiv \delta Q - \widehat{\delta Q} \hspace{1.2cm}
{\rm with} \hspace{1cm} Q = Z_B, Z_W, g^2, \; {\rm etc...}
\end{equation}
These equations give the differences among the renormalization
constants of the SM in the large $\mh$ limit and those in the
EChL, when the on-shell renormalization scheme is chosen in
both theories. They are obtained here as a constraint imposed
by the matching; one can also calculate them from the explicit
expressions of the on-shell counterterms of the two theories
and verify that these relations are indeed satisfied.

\section{Renormalization of the Effective Theory}

In this section we briefly describe the renormalization procedure in
the effective theory.  Given the effective Lagrangian of
Eq.(\ref{ECL}), the first step is to redefine the fields and
parameters of the Lagrangian according to Eq.(\ref{RET}).  It
introduces, at a formal level, the set of counterterms of the
effective theory $\widehat{\delta Z}_i, \widehat{\delta g}$, etc,
that need to be computed once a particular renormalization
prescription scheme is chosen. We fix here the on-shell
renormalization scheme as we did in the case of the SM. For practical
reasons we prefer to choose the renormalization conditions as in
reference \cite{HO}, which are the most commonly used for LEP
physics. In terms of the renormalized selfenergies these
renormalization conditions read as follows
\begin{equation}
\widehat{\Sigma}^{\rm R}_{W} (M_W^2) = 0,\hspace{3mm}
\widehat{\Sigma}^{\rm R}_{Z} (M_Z^2) = 0,\hspace{3mm}
\widehat{\Sigma}^{\prime \; \rm R}_{\gamma} (0) = 0,\hspace{3mm}
\widehat{\Sigma}^{\rm R}_{\gamma Z} (0) = 0. \label{RCE}
\end{equation}
The renormalized self energies are computed in the effective theory
as usual, namely, by adding all the contributions from the one
loop diagrams and from the counterterms. We get the following
expressions\footnote{Notice that in our rotation defining the physical
gauge fields, the terms in $s$ have different sign than in reference
\cite{HO}.}:
\begin{eqnarray}
\widehat{\Sigma}^{\rm R}_{\gamma} (q^2) & = &
\widehat{\Sigma}^{\rm L}_{\gamma} (q^2) +
\left( s^2 \widehat{\delta Z_W} + c^2 \widehat{\delta Z_B}
\right) q^2 + s^2 g^2 (a^b_8 - 2 a^b_1) q^2.
\nonumber\\[2mm]
\widehat{\Sigma}^{\rm R}_{W} (q^2) & = &
\widehat{\Sigma}^{\rm L}_{W} (q^2) +
\widehat{\delta Z_W} \left( q^2 - M_W^2 \right)
- \widehat{\delta M_W^2}. \nonumber\\[2mm]
\widehat{\Sigma}^{\rm R}_{Z} (q^2) & = &
\widehat{\Sigma}^{\rm L}_{Z} (q^2) +
\left( c^2 \widehat{\delta Z_W} + s^2 \widehat{\delta Z_B}
\right) ( q^2 - M_Z^2) - \widehat{\delta M_Z^2} \nonumber\\
& &
+ 2 g'^2 a_0^b M_Z^2 + \left( 2 s^2 g^2  a^b_1 + c^2 g^2 a_8^b
 + (g^2 + g^{\prime 2}) a^b_{13} \right) q^2.
\nonumber\\[2mm]
\widehat{\Sigma}^{\rm R}_{\gamma Z} (q^2) & = &
\widehat{\Sigma}^{\rm L}_{\gamma Z} (q^2) +
s c \left( \widehat{\delta Z_W} - \widehat{\delta Z_B}
\right) q^2 - s c\; M_Z^2 \left( \frac{\widehat{\delta g'}}{g'} -
\frac{\widehat{\delta g}}{g} \right) \nonumber\\
& & + \left( s c g^2 a^b_8 - (c^2 - s^2) g g' a^b_1 \right) q^2.
\label{JQL}
\end{eqnarray}
where
\begin{eqnarray}
\widehat{\delta M_W^2} & = & M_W^2 \left(
\widehat{\delta Z_\Phi} - 2 \frac{\widehat{\delta g}}{g}
- 2 \frac{\widehat{\delta v}}{v} - \widehat{\delta Z_W}
\right), \nonumber\\[2mm]
\widehat{\delta M_Z^2} & = & M_Z^2 \left(
\widehat{\delta Z_\Phi} - 2 c^2 \frac{\widehat{\delta g}}{g}
- 2 s^2 \frac{\widehat{\delta g'}}{g'}
- 2 \frac{\widehat{\delta v}}{v} -
c^2 \widehat{\delta Z_W} - s^2 \widehat{\delta Z_B}
\right),\nonumber\\[2mm]
M_W^2 & = & g^2 v^2 / 4, \nonumber\\[2mm]
M_Z^2 & = & (g^2 + g'^2) v^2 / 4,
\label{masas}
\end{eqnarray}
and the superscripts R and L denote renormalized and EChL
loops respectively.

{}From Eq.(\ref{masas}) the following relation among the
$W$ and $Z$ mass counterterms is obtained
\begin{equation}
\frac{\widehat{\delta M_Z^2}}{M_Z^2} -
\frac{\widehat{\delta M_W^2}}{M_W^2} =
 2 s^2 \frac{\widehat{\delta g}}{g} +
 2 c^2 \frac{\widehat{\delta g'}}{g'}
+ s^2 \left(\widehat{\delta Z_W} - \widehat{\delta Z_B}
\right)
\end{equation}
Finally, by requiring these renormalized self energies to fulfill
Eq.(\ref{RCE}) and taking into account that the $U(1)_{\rm Y}$ Ward
identity implies $\widehat{\delta g'} = 0$ one gets the following
results for the values of the counterterms in terms of the
unrenormalized selfenergies of the effective theory and the bare
$a_i$'s:
\begin{eqnarray}
\widehat{\delta M_W^2} & = &
\widehat{\Sigma}^{\rm L}_W (M_W^2), \nonumber\\[2mm]
\widehat{\delta M_Z^2} & = &
\widehat{\Sigma}^{\rm L}_{Z} (M_Z^2) + M_Z^2
\left( 2 g'^2 a_0^b + 2 s^2 g^2  a^b_1 + c^2 g^2 a_8^b
 + (g^2 + g^{\prime 2}) a^b_{13} \right), \nonumber\\[2mm]
\frac{\widehat{\delta g}}{g} & = &
\frac{- 1}{s c} \frac{\widehat{\Sigma}^{\rm L}_{ \gamma Z}(0)}
{M_Z^2}, \nonumber\\[2mm]
\frac{\widehat{\delta g'}}{g'} & = & 0, \nonumber\\[2mm]
\widehat{\delta Z_W} & = &
\frac{c^2}{s^2} \left(\frac{\widehat{\Sigma}^{\rm L}_{Z}(M_Z^2)}
{M_Z^2} - \frac{\widehat{\Sigma}^{\rm L}_{W}(M_W^2)}
{M_W^2} \right) + 2 \frac{c}{s} \frac{
\widehat{\Sigma}^{\rm L}_{ \gamma Z}(0)}{M_Z^2} -
\widehat{\Sigma}^{\prime \;\rm L}_\gamma (0) \nonumber\\
& & + 2 g^2 a^b_0 + 2 g^2 a^b_1 +
\frac{c^2 - s^2}{s^2} g^2 a^b_8 + \frac{c^2}{s^2}
(g^2 + g'^2) a^b_{13}, \nonumber\\[2mm]
\widehat{\delta Z_B} & = &
 \frac{\widehat{\Sigma}^{\rm L}_{W}(M_W^2)}{M_W^2} -
\frac{ \widehat{\Sigma}^{\rm L}_{Z}(M_Z^2)}{M_Z^2} -
2 \frac{s}{c} \frac{\widehat{\Sigma}^{\rm L}_{ \gamma Z}(0)}{M_Z^2} -
\widehat{\Sigma}^{\prime \rm L}_\gamma (0) \nonumber\\
& & - \left( 2 g'^2 a^b_0 + g^2 a^b_8 +
(g^2 + g'^2) a^b_{13} \right). \label{ECT}
\end{eqnarray}

Now that we have at hand Eq.(\ref{ECT}) the only parameters of the
theory that still need to be renormalized are the electroweak chiral
parameters $a_i$. The following formal redefinition of the chiral
parameters has already been introduced
\begin{equation}
a^b_i = a_i (\mu) + \delta a_i. \label{NSQ}
\end{equation}
The divergent part of the $a_i^b$ parameters, or equivalently the
divergent part of the counterterms $\delta a_i$, are fixed by
the symmetries of the effective theory and since the work of
Longhitano \cite{LON} they are known to be
\begin{eqnarray}
& & \delta a_0 |_{div} = \frac{1}{16 \pi^2} \frac{3}{8} \Delta_\epsilon ,
\hspace{2cm}\delta a_1 |_{div} = \frac{1}{16 \pi^2} \frac{1}{12}
\Delta_\epsilon , \nonumber \\[2mm]
& & \delta a_2 |_{div} = \frac{1}{16 \pi^2} \frac{1}{24} \Delta_\epsilon ,
\hspace{2cm}\delta a_3 |_{div} = \frac{- 1}{16 \pi^2} \frac{1}{24}
\Delta_\epsilon , \nonumber \\[2mm]
& & \delta a_4 |_{div} = \frac{- 1}{16 \pi^2} \frac{1}{12} \Delta_\epsilon ,
\hspace{2cm}\delta a_5 |_{div} = \frac{- 1}{16 \pi^2} \frac{1}{24}
\Delta_\epsilon , \nonumber \\[2mm]
& & \delta a_i |_{div} = 0, \hspace{5mm} i = 6,....13. \label{PP}
\end{eqnarray}
These universal divergent contributions to the chiral bare
parameters imply in turn the already mentioned universal
scale dependence of the renormalized parameters
\begin{eqnarray}
& &a_0(\mu)  =  a_0(\mu ') + \frac{1}{16 \pi^2}
\frac{3}{8} \log\frac{\mu^2}{\mu '^2}, \hspace{1cm}
a_1(\mu)  =  a_1(\mu ') + \frac{1}{16 \pi^2}
\frac{1}{12} \log\frac{\mu^2}{\mu '^2}, \nonumber\\[2mm]
& &a_2(\mu)  =  a_2(\mu ') + \frac{1}{16 \pi^2}
\frac{1}{24} \log\frac{\mu^2}{\mu '^2}, \hspace{1cm}
a_3(\mu)  =  a_3(\mu ') - \frac{1}{16 \pi^2}
\frac{1}{24} \log\frac{\mu^2}{\mu '^2}, \nonumber\\[2mm]
& &a_4(\mu)  =  a_4(\mu ') - \frac{1}{16 \pi^2}
\frac{1}{12} \log\frac{\mu^2}{\mu '^2}, \hspace{1cm}
a_5(\mu)  =  a_5(\mu ') - \frac{1}{16 \pi^2}
\frac{1}{24} \log\frac{\mu^2}{\mu '^2}, \nonumber\\[2mm]
& &a_i(\mu) = a_i(\mu') ; \hspace{3mm} i = 6,...13.
\end{eqnarray}

The value of the bare chiral parameters $a^b_i$, on the other hand,
is completely determined by the matching procedure in terms of the
renormalized parameters of the underlying physics that has been
integrated out, as we have seen for the particular case of a heavy
Higgs.  However, for a given $a^b_i$, we still have to choose how to
separate the finite part into the renormalized $a_i(\mu)$ and the
counterterm $\delta a_i$ in Eq.(\ref{NSQ}) such that their sum gives
$a^b_i$.  This second renormalization scheme concerns only to the
effective theory.  Therefore, in using a set of renormalized
parameters $a_i(\mu)$ for a particular underlying theory, one must
specify, in addition, how the finite parts of the counterterms in
Eq.(\ref{NSQ}) have been fixed.

In the case of the SM, where a heavy Higgs has been integrated
out to one loop, the bare chiral parameters are given in
Eq.(\ref{aMH}). They correspond to the on-shell renormalization
of the underlying SM. Now, in order to present the corresponding
renormalized parameters we have to fix the finite parts of the
counterterms. For instance, if we fix the counterterms to include
just the $\Delta_\epsilon$ terms as in Eq.(\ref{PP}),
the renormalized chiral parameters for the SM with a heavy Higgs
are\footnote{This particular renormalization of the chiral parameters
was chosen in our first work\cite{HR2}, where we called it $\msb$.}:
\begin{equation}
\begin{array}{ll}
a_0 (\mu)  =  {\displaystyle  \frac{1}{16 \pi^2}
\frac{3}{8}\left( \frac{5}{6} - \log\frac{M_H^2}{\mu^2} \right)},&
\hspace{-1cm}a_3 (\mu) =  {\displaystyle \frac{-1}{16 \pi^2}
\frac{1}{24}
\left( \frac{17}{6} - \log\frac{M_H^2}{\mu^2} \right)},\\[6mm]
a_1 (\mu)  =  {\displaystyle \frac{1}{16 \pi^2}  \frac{1}{12}
\left( \frac{5}{6} - \log\frac{M_H^2}{\mu^2} \right)}, &
\hspace{-1cm}a_4 (\mu) =  {\displaystyle \frac{- 1}{16 \pi^2}
\frac{1}{12}
\left( \frac{17}{6} - \log\frac{M_H^2}{\mu^2} \right)},\\[6mm]
a_2 (\mu)  =  {\displaystyle \frac{1}{16 \pi^2} \frac{1}{24}
\left( \frac{17}{6} - \log\frac{M_H^2}{\mu^2} \right)}, &
\hspace{-1cm}a_{11} (\mu)  = {\displaystyle
\frac{-1}{16 \pi^2}\frac{1}{24}},\\[6mm]
a_5 (\mu)  =  {\displaystyle \frac{v^2}{8 \mh^2} -
 \frac{1}{16 \pi^2}  \frac{1}{24}
\left( \frac{79}{3} - \frac{ 27 \pi}{2 \sqrt{3}}
 - \log\frac{M_H^2}{\mu^2} \right)}, \\[6mm]
a_i (\mu) = 0, \hspace{3mm} i = 6,7,8,9,10,12,13.
\end{array} \label{aR}
\end{equation}

We have discussed\cite{HR3} also other choices of the renormalization
scheme of the effective theory, for instance the one chosen by Gasser
and Leutwyler for the linear $O(N)$  sigma model\cite{GL1} .

\section{Calculating Observables with the EChL}

In this section we will show, as an example, the explicit calculation
of the radiative corrections to $\Delta \rho$ and $\Delta r$ within
the electroweak chiral Lagrangian approach. These observables are
defined in the effective theory in terms of the renormalized
self-energies in the same way as in the fundamental SM, namely:
\begin{eqnarray}
\Delta \rho & \equiv & \frac{\widehat{\Sigma}^{\rm R}_Z (0)}{M_Z^2}
-  \frac{\widehat{\Sigma}^{\rm R}_W (0)}{M_W^2}, \nonumber\\[2mm]
\Delta r & \equiv &  \frac{\widehat{\Sigma}^{\rm R}_W (0)}{M_W^2}
+ {\rm (vertex + box)}, \label{ROR}
\end{eqnarray}
where
\begin{displaymath}
{\rm (vertex + box)} \equiv \frac{g^2}{16 \pi^2} \left(
6 + \frac{ 7 - 4 s^2}{2 s^2} \log c^2 \right)
\end{displaymath}
and the renormalized self-energies can be computed as we have
explained in section 6.

Once a renormalization scheme has been chosen, one can always express
$\Delta \rho$ and $\Delta r$ in terms of unrenormalized self-energies
and the $a_i^b$'s. For instance, in the on-shell scheme
given by the conditions of Eq.(\ref{RCE}), one gets the particular
values of the counterterms given in Eq.(\ref{ECT}). Next, by plugging
these counterterms into Eq.(\ref{JQL}), one obtains the renormalized
self-energies in terms of the unrenormalized ones and the
$a_i^b$'s. Finally, by substituing these formulas into
Eq.(\ref{ROR}) the following expressions for $\Delta \rho$ and
$ \Delta r$ in the on-shell scheme are found
\begin{eqnarray}
\Delta \rho & = & \frac{\widehat{\Sigma}^{\rm L}_Z (0)}{M_Z^2}
-  \frac{\widehat{\Sigma}^{\rm L}_W (0)}{M_W^2}
+ \frac{2 s}{c} \frac{\widehat{\Sigma}^{\rm L}_{\gamma Z} (0)}{M_Z^2}
+ 2 g'^2 a^b_0, \nonumber \\[2mm]
\Delta r & = & \frac{\widehat{\Sigma}^{\rm L}_W (0) -
\widehat{\Sigma}^{\rm L}_W (M_W^2)}{M_W^2} +
\widehat{\Sigma}'^{\rm  L}_\gamma (0) + \frac{c^2}{s^2}
\left[ \frac{\widehat{\Sigma}^{\rm L}_W (M_W^2)}{M_W^2}
- \frac{\widehat{\Sigma}^{\rm L}_Z (M_Z^2)}{M_Z^2}
-  \frac{2 s}{c} \frac{\widehat{\Sigma}^{\rm L}_{\gamma Z} (0)}
{M_Z^2} \right] \nonumber\\[2mm]
& & - 2 g^2 a^b_0 + \frac{s^2 - c^2}{s^2}
g^2  (a^b_8 + a^b_{13} ) - 2 g^2 (a^b_1 + a^b_{13}) +
 {\rm (vertex + box)}.
\end{eqnarray}

The explicit computation of the bosonic loop contributions in the
effective theory, as well as the contributions from just
$a^b_0$ and $a^b_1$ were found in \cite{DEH,EH}.
We present here the complete result
\begin{eqnarray}
\Delta \rho & = & \frac{g^2}{16 \pi^2} \left[
 \frac{3}{4} \frac{s^2}{c^2} \left( - \Delta_\epsilon
+ \log \frac{M_W^2}{\mu^2} \right) + h(M_W^2,M_Z^2)
 \right]
+ 2 g'^2 a^b_0, \nonumber \\[2mm]
\Delta r & = & \frac{g^2}{16 \pi^2} \left[
\frac{11}{12} \left( \Delta_\epsilon - \log \frac{M_W^2}{\mu^2}
\right) + f(M_W^2,M_Z^2) \right] \nonumber\\[2mm]
& & - 2 g^2 a^b_0 + \frac{s^2 - c^2}{s^2}
g^2  (a^b_8 + a^b_{13} ) - 2 g^2 (a^b_1 + a^b_{13}). \label{RORAB}
\end{eqnarray}
where
\begin{eqnarray}
h(M_W^2,M_Z^2) & = &  \frac{1}{c^2} \log c^2 \left( \frac{17}{4 s^2} -
7 + 2 s^2 \right) + \frac{17}{4} - \frac{5}{8} \frac{s^2}{c^2}
\nonumber\\[2mm]
f(M_W^2,M_Z^2) & = & \log c^2 \left( \frac{5}{c^2} -1 +
\frac{3 c^2}{s^2} - \frac{17}{4 s^2 c^2} \right)
- s^2 (3 + 4 c^2) F(M_Z^2,M_W,M_W)\nonumber \\
& & + I_2(c^2)(1 - \frac{c^2}{s^2}) + \frac{c^2}{s^2} I_1(c^2) +
\frac{1}{8 c^2} ( 43 s^2 - 38) \nonumber \\
& & + \frac{1}{18} (154 s^2 - 166 c^2) + \frac{1}{4 c^2} +
\frac{1}{6} + \Delta \alpha +
\left( 6 + \frac{7 - 4 s^2} {2 s^2} \log c^2 \right).
\end{eqnarray}
and $F$, $I_1$, $I_2$ and $\Delta \alpha$ can be found in \cite{MS}.
In Eq.(\ref{RORAB}) there are apparently a divergent term and a $\mu$-scale
dependent term. However, when one redefines the bare effective chiral
parameters as usual, $a^b_i = a_i(\mu) + \delta a_i$, it can be easily
seen that the divergent terms are cancelled by the divergent parts
of the $\delta a_i$ and the $\mu$-scale dependence is cancelled by the
scale dependence of the $a_i(\mu)$. The observables $\Delta \rho$ and
$\Delta r$ turn out to be finite and scale and renormalization
prescription independent, as it must be. In particular, if we set the
substraction scheme for the chiral counterterms
to include just the $\Delta_\epsilon$ terms as in Eq.(\ref{PP}),
the following expressions for the bosonic contributions to
$\Delta \rho$ and $\Delta r$
in terms of renormalized chiral parameters are obtained
\begin{eqnarray}
\Delta \rho & = & \frac{g^2}{16 \pi^2} \left[
 \frac{3}{4} \frac{s^2}{c^2} \log \frac{M_W^2}{\mu^2}
+ h(M_W^2, M_Z^2) \right]
+ 2 g'^2 a_0(\mu), \nonumber \\[2mm]
\Delta r & = & \frac{g^2}{16 \pi^2} \left[
 - \frac{11}{12}
\log \frac{M_W^2}{\mu^2} + f(M_W^2,M_Z^2) \right] \nonumber\\[2mm]
& & - 2 g^2 a_0(\mu) + \frac{s^2 - c^2}{s^2}
g^2  (a_8 + a_{13} ) - 2 g^2 (a_1(\mu) + a_{13}). \label{RRF}
\end{eqnarray}

Equations (\ref{RRF}) are general and can be applied to any
underlying physics for the symmetry breaking sector.
If we want to recover the values of $\Delta \rho$ and $\Delta r$
in the particular case of the SM with a heavy Higgs,
one just has to substitute the values of the chiral parameters
in Eq.(\ref{aR}) into Eq.(\ref{RRF}) to obtain
\begin{eqnarray}
\Delta \rho & = & \frac{g^2}{16 \pi^2} \left[
 - \frac{3}{4} \frac{s^2}{c^2} \left(
\log \frac{M_H^2}{M_W^2} - \frac{5}{6} \right)
+ h(M_W^2, M_Z^2) \right], \nonumber \\[2mm]
\Delta r & = & \frac{g^2}{16 \pi^2} \left[
 \frac{11}{12}
\left(\log \frac{M_H^2}{M_W^2} - \frac{5}{6} \right)
 + f(M_W^2,M_Z^2) \right].
\end{eqnarray}
which agrees with the result given in \cite{MS}.

One can similarly obtain the heavy Higgs contributions to other
relevant observables in electroweak phenomenology.

\section*{Acknowledgements}

We thank the organizers of the workshop for their invitation
and all the members of the theory group at E\H{o}tv\H{o}s University
for their kind hospitality.


\begin{thebibliography}{99}
\bibitem{DH} A. Dobado and M.J. Herrero, {\it Phys. Lett.} {\bf B228}
(1989),495; {\bf B233} (1989),505.\\
J. Donoghue and C. Ramirez, {\it Phys. Lett.} {\bf B234} (1990), 361.
\bibitem{FER} For a recent review on effective
chiral lagrangians in electroweak physics see \\
F. Feruglio, {\it Int. J. Mod. Phys.} {\bf A8} (1993), 4937.
\bibitem{AB} T.Appelquist and C.Bernard, {\it Phys. Rev.} {\bf D22}
(1980), 200;\\
T.Appelquist, in {\it Gauge Theories and Experiments at High
Energies} (Ed. K.C.Browner and D.G.Sutherland, Scottish U.
Summer School, 1980).
\bibitem{LON} A.C.Longhitano, {\it Nucl. Phys.} {\bf B188} (1981), 118;
 {\it Phys. Rev.} {\bf D22} (1980), 1166.
\bibitem{HR2} M.J. Herrero and E. Ruiz Morales,
 {\it Nucl. Phys.} {\bf B418} (1994), 431.
\bibitem{HR3} M.J. Herrero and E. Ruiz Morales, preprint FTAUM94/11
and hep-ph/9411207.
\bibitem{W} S. Weinberg, {\it Physica} {\bf 96A} (1979), 327.
\bibitem{GL1} J. Gasser and H. Leutwyler, {\it Ann. Phys. (N.Y.)}
 {\bf 158} (1984), 142;
\bibitem{BDV}  S. Dawson and G. Valencia, {\it Nucl. Phys.}
{\bf B352} (1991), 27.\\
J. Barger, S. Dawson and G. Valencia, {\it Nucl. Phys.}
{\bf 399} (1993), 364.
\bibitem{DHT} A. Dobado, M.J. Herrero and J. Terron, {\it Z. Phys.}
{\bf C50} (1991), 205; {\it Z. Phys.} {\bf C50} (1991), 465.
\bibitem{HR1} M.J. Herrero and E. Ruiz Morales, {\it Phys. Lett.}
 {\bf B296} (1992), 397.
\bibitem{MAY} A. Dobado and M.T. Urdiales, preprint FTUAM94/29 and
 FTUCM94/19.
\bibitem{HT} B. Holdom and J. Terning, {\it Phys. Lett.} {\bf B247}
(1990), 88.
\bibitem{DEH} A. Dobado, D. Espriu and M.J. Herrero, {\it Phys. Lett.}
{\bf B255} (1991), 405.
\bibitem{EH} D. Espriu and M.J. Herrero, {\it Nucl. Phys.} {\bf B373}
(1992), 117.
\bibitem{PES} B.W. Lynn, M.E. Peskin and R.G. Stuart, CERN-86-02
(1986).\\
M.E. Peskin and T. Takeuchi, {\it Phys. Rev. Lett.} {\bf 65}
(1990), 964; {\it Phys. Rev.} {\bf D46} (1992), 381. \\
M. Golden and L. Randall, {\it Nucl. Phys.} {\bf B361} (1991), 3.
\bibitem{GEO} H. Georgi, {\it Ann. Rev. Nucl. Part. Sci.}
{\bf 43} (1993), 209 and references therein.
\bibitem{DOM} D. Espriu and J. Matias, preprint UB-ECM-PR-94/12.
\bibitem{SANTA} M. Bilenky and A. Santamar\'{\i}a,
 {\it Nucl. Phys.} {\bf B420} (1994), 47.
\bibitem{MAW} W. Marciano and S. Willenbrock, {\it Phys. Rev.}
{\bf D40} (1988), 2509.
\bibitem{DW} S. Dawson and S. Willenbrock, {\it Phys. Rev.}
{\bf D40} (1989), 2880.
\bibitem{EQV} The equivalence theorem in effective lagrangians
has been discussed in \\A. Dobado and J.R. Pel\'aez, {\it Nucl. Phys.}
{\bf B425} (1994), 110. \\
H.-Y. He, Y.-P. Kuang and X. Li, {\it Phys. Lett.} {\bf B329} (1994),
278.
\bibitem{HO} M. Bohm, H. Spiesberger and W. Hollik,
 {\it Fortschr. Phys.} {\bf 34}, 11, (1986), 687.
\bibitem{MS} W. Marciano and A. Sirling, {\it Phys. Rev.} {\bf D22}
(1980), 2697. \\
G. Burgers and W. Hollik in {\it Polarization at LEP}
(CERN Yellow Report, \\ed. G. Alexander et al., CERN, Geneva, 1989).\\
M. Consoli and W. Hollik; G. Burgers and F. Jegerlehner in
 {\it Z Physics at LEP} \\(CERN Yellow Report. ed. G. Altarelli et al.,
CERN, Geneva, 1989).
\end{thebibliography}
\end{document}